\documentclass[twocolumn, floatfix]{aastex631}
\usepackage{tikz,amsmath}
\usepackage{color,soul,blindtext}
\usepackage[english]{babel}
\usepackage{fancyvrb,float,hyperref}
\hypersetup{linkcolor=red,citecolor=cyan,filecolor=magenta,urlcolor=blue}
\usepackage{natbib}

\newcommand{\Msol}{M$_{\odot}$}

\newcommand{\lsim}{$\lesssim$}
\newcommand{\gsim}{$\gtrsim$}
\newcommand{\galfit}{\textsc{Galfit}}
\newcommand{\zrange}{$0.5<z<1.5$}
\begin{document}

\title{\bf Recent star formation in $0.5<z<1.5$ quiescent galaxies}

\author[0000-0001-7016-5220]{Michael~J.~Rutkowski}
\affiliation{Minnesota State University-Mankato, Dept. of Physics \& Astronomy, Trafton Science Center North 141, Mankato, MN, 56001 USA}
\email{michael.rutkowski@mnsu.edu}

\author[0000-0002-7830-363X]{Bonnabelle Zabelle}
\affiliation{Minnesota Institute for Astrophysics, University of Minnesota, 116 Church St SE, Minneapolis, MN 55455, USA}

\author[1234-5678-8765-4321]{Tyler Hagen}
\affiliation{University of Utah, Dept. of Physics \& Astronomy, 115 S. 1400 E., Salt Lake City, UT 84112-0830, USA}

\author[0000-0003-3329-1337]{Seth Cohen} 
\affiliation{School of Earth and Space Exploration, Arizona State University, Tempe, AZ 85287, USA}

\author[1234-5678-8765-4321]{Christopher Conselice}
\affiliation{School of Physics and Astronomy, The University of Nottingham, University Park, Nottingham NG7 2RD, UK}

\author[0000-0001-9440-8872]{Norman Grogin}
\affiliation{Space Telescope Science Institute, 3700 San Martin Drive, Baltimore, MD 21218, USA}

\author[0000-0003-2775-2002]{Yicheng Guo}
\affiliation{Department of Physics and Astronomy, University of Missouri, Columbia, MO 65211, USA}

\author[0000-0001-8587-218X]{Matthew Hayes}
\affiliation{Stockholm University, Department of Astronomy and Oskar Klein Centre for Cosmoparticle Physics, SE-10691, Stockholm, Sweden}

\author[0000-0002-5601-575X]{Sugata Kaviraj}
\affiliation{Centre for Astrophysics Research, University of Hertfordshire, Hatfield, AL10 9AB, UK}

\author[1234-5678-8765-4321]{Anton Koekemoer}
\affiliation{Space Telescope Science Institute, 3700 San Martin Drive, Baltimore, MD 21218, USA}

\author[0000-0003-1581-7825]{Ray A. Lucas}
\affiliation{Space Telescope Science Institute, 3700 San Martin Drive, Baltimore, MD 21218, USA}

\author[0000-0002-6016-300X]{Kameswara Bharadwaj Mantha}
\affiliation{Minnesota Institute for Astrophysics, University of Minnesota, 116 Church St SE, Minneapolis, MN 55455, USA}

\author[0000-0002-6632-4046]{Alec Martin}
\affiliation{Department of Physics and Astronomy, University of Missouri, Columbia, MO 65211, USA}

\author[0000-0001-7166-6035]{Vihang Mehta}
\affiliation{IPAC, California Institute of Technology, 1200 E. California Blvd., Pasadena CA, 91125, USA} 

\author[0000-0001-5846-4404]{Bahram Mobasher}
\affiliation{Department of Physics and Astronomy, University of California, Riverside, Riverside, CA 92521, USA}

\author[0000-0001-6145-5090]{Nimish Hathi}
\affiliation{Space Telescope Science Institute, 3700 San Martin Drive, Baltimore, MD 21218, USA}

\author[0000-0001-5294-8002]{Kalina~V.~Nedkova}
\affiliation{Department of Physics and Astronomy, Johns Hopkins University, Baltimore, MD 21218, USA}
\affiliation{Space Telescope Science Institute, 3700 San Martin Drive, Baltimore, MD 21218}

\author[0000-0002-8190-7573]{Robert O'Connell}
\affiliation{Department of Astronomy, University of Virginia, Charlottesville, VA 22904}

\author[0000-0002-9946-4731]{Marc Rafelski}
\affiliation{Space Telescope Science Institute, 3700 San Martin Drive, Baltimore, MD 21218, USA}

\author[0000-0002-9136-8876]{Claudia Scarlata}
\affiliation{Minnesota Institute for Astrophysics, University of Minnesota, 116 Church St SE, Minneapolis, MN 55455, USA}

\author[0000-0002-7064-5424]{Harry~I.~Teplitz}
\affiliation{IPAC, California Institute of Technology, 1200 E. California Blvd., Pasadena CA, 91125, USA} 

\author[0000-0002-9373-3865]{Xin Wang}
\affiliation{School of Astronomy and Space Science, University of Chinese Academy of Sciences (UCAS), Beijing 100049, China}
\affiliation{National Astronomical Observatories, Chinese Academy of Sciences, Beijing 100101, China}
\affiliation{Institute for Frontiers in Astronomy and Astrophysics, Beijing Normal University, Beijing 102206, China}

\author[0000-0001-8156-6281]{Rogier Windhorst}
\affiliation{School of Earth and Space Exploration, Arizona State University, Tempe, AZ 85287, USA}

%
\author[0000-0003-3466-035X]{{L. Y. Aaron} {Yung}}
\affiliation{Space Telescope Science Institute, 3700 San Martin Dr., Baltimore, MD 21218, USA}

\author{the UVCANDELS Team}

\date{\today}


\begin{abstract}
Observations of massive, quiescent galaxies reveal a relatively uniform evolution: following prolific star formation in the early universe, these galaxies quench and transition to their characteristic quiescent state in the local universe. The debate on the relative role and frequency of the process(es) driving this evolution is robust. In this letter, we identify $0.5 \lesssim z \lesssim 1.5$ massive, quiescent galaxies in the HST/UVCANDELS extragalactic deep fields using traditional color selection methods and model their spectral energy distributions, which incorporates novel UV images. This analysis reveals $\sim15\%$ of massive, quiescent galaxies have experienced minor, recent star formation($<10\%$ of total stellar mass within the past $\sim$1Gyr). We find only a marginal, positive correlation between the probability for recent star formation and a measure of the richness of the local environment from a statistical analysis. Assuming the recent star formation present in these quiescent galaxies is physically linked to the local environment, these results suggest only a minor role for dynamic external processes (galaxy mergers and interactions) in the formation and evolution of these galaxies at this redshift.\vspace{-15pt}
\end{abstract} \vspace{-60pt}

\section{Introduction}\label{sec:intro}  
The development of a robust model of the transformation of high-redshift star-forming galaxies into quiescent galaxies observed in the local universe is a primary goal in the study of galaxy evolution. Massive quiescent galaxies, specifically, are an appealing class for study in the development of such models. Historically, diverse lines of observational evidence---e.g., optical colors, inferred star formation histories, and $\alpha$-element enhancement \citep[e.g.,][respectively]{Bower1992,Heavens2004,Johannson2012} --- have been interpreted as evidence for a common evolutionary history, whereby quiescent galaxies descended from galaxies in the early universe that rapidly assembled, ceased star formation (quench), and evolve quiescently until the present day. 

Extensive surveys have measured the declining quiescent fraction (with a commensurate increase in stellar mass density arising from the quenching of galaxies) with increasing redshift, since $z\lesssim4$ \citep{Muzzin2013,Ilbert2013}. More recently, massive quiescent galaxies have been identified in the first $\sim2$ Gyr \citep[at $z\sim$4,][]{Glazebrook2017,Valentino2020} that have recently quenched intense star formation.  JWST is dramatically advancing the state of the art, detecting similar mass, quenched galaxies at higher redshifts \citep[$z>4$,][]{Carnall23a,Carnall23,deGraaff2024} while also revealing possible progenitors ($z\gg5$, low-mass, optically galaxies) effectively ``caught in the act'', transitioning to quiescence through (mini)-quenching events \citep{Strait2023,Looser2024} with the clear spectroscopic signatures of the characteristic rapid
star formation, quenching, and declining star formation rate thereafter. In cosmological simulations, a combination of “in-situ” \citep[stellar and AGN feedback;][]{Dubois2013,
El-Badry2016} and 
“external” \citep[i.e., environmental;][]{Kaviraj2015} processes are typically invoked to broadly reproduce the evolution of this class of galaxies towards the quiescent state. 

But, refining this general evolutionary scenario is necessary to accommodate a variety of key observational caveats.  First, a substantial fraction (exceeding $\sim20\%$ at $z\lesssim2\%$) of quenched or quiescent galaxies exhibit recent star formation\citep{Kaviraj2007,Kaviraj2008,Rutkowski2014,Kim2018,Paspaliaris2023}, and may do so repeatedly, via mini-quenching events which only temporarily halt star formation \citep{Dome2024,Gelli2025}.  Depending on the epoch in which these galaxies are observed during this process, the inferred star formation history (SFH) for such galaxies may not exhibit the form characteristic of these galaxies---a uniform single exponentially-declining SFH with star formation rate, $SFR\propto\exp\left(-t/\tau\right)$---that is typically used for pre-selection \citep[on specific SFR, e.g.,][]{Salim2018}. 

Secondly, the relative predominance of the varied physical processes responsible for the transformation of quiescent galaxies over time is uncertain. Considering only mergers, their impact is diverse and varied. Recognizing that recent observations indicate a likely mass-dependent effect \citep{Cutler23}, recent star formation in quiescent galaxies is likely a result of mergers or accretion \citep{Kaviraj2013,Cleland2021}. Further, the episodic accretion of low-mass satellites could \citep[in theory, see][]{Naab2009} motivate their observed size-mass growth \citep{Cassata2011,Ryan2012}. In contrast, major mergers are implicated in the rapid quenching of star formation \citep{Bekki2005, Verrico2023} in galaxies which subsequently manifest as quiescent, post-starburst (PSBs) galaxies \citep[which are observed preferentially in richer environments,][]{Poggianti2009}.

Large surveys of recent star formation (RSF) in quiescent galaxies can provide insight to these complex, stochastic mechanisms and the frequency and duration of the quenching process \citep{Wild2016, Belfiore2018, Clausen2024}. In particular, when rest-frame UV-optical-nearIR photometry are available for quiescent galaxies, it is possible to uniquely characterize the age, mass fraction, and star formation history (SFH) of RSF.

Only HST WFC3/UVIS can provide high ($\ll 1\farcs$) resolution\footnote{AstroSat UVIT/N242W can image
at comparable wavelengths as HST UVIS F275W, but at $\sim5\times$ lower spatial resolution}
rest-frame UV imaging of \zrange\,galaxies, but prior to HST Cycle 26, only $\sim100$ sq.arcmin {\it in total} had broadband rest-frame far-UV imaging {\it and} the longer wavelength imaging necessary for characterizing the predominant, old stellar population.

In this Letter, we directly search for RSF in quiescent galaxies in new UV extragalactic survey data. Throughout this letter we assume a Planck concordance model \citep{Planck2016}; we report magnitudes as AB magnitudes \citep{Oke1983}.\\ \vspace*{-12pt}

\section{UVCANDELS Imaging and Photometry}\label{subsec:uvcdata} 

UVCANDELS obtained WFC3/UVIS F275W and ACS F435W images over $\sim$450 sq.~arcmin. in four extragalactic deep fields (COSMOS, EGS, GOODS-North and GOODS-South) to a depth of AB=27 (3$\sigma$ Point Source). To date, this UV imaging has already supported diverse analyses in Lyman Continuum escape \citep{Smith2024}, resolved star formation \citep{Mehta2023, Nedkova24}, and quiescent galaxies \citep{Zabelle2023}. Combined with the rich multiwavelength imaging and spectroscopy available for these fields, UVCANDELS effectively {\it quadruples} the area in which RSF can be directly probed in $z\sim1$ massive, quiescent galaxies as F275W and F435W are sensitive to  $1600\lesssim\lambda$[\AA]$\lesssim2000$ \& $2500\lesssim\lambda$[\AA]$\lesssim3200$ ($960\lesssim\lambda$[\AA]$\lesssim1200$ \& $1500\lesssim\lambda$[\AA]$\lesssim2000$) at $z\sim0.5$ ($z\sim1.5$).\\ \vspace*{-15pt}

We refer the reader to \cite{Wang2025} and \cite{Mehta2024} for full details, but reiterate key components of the data processing here.  Namely, custom routines produced by \cite{Rafelski2015} and \cite{Prichard2022} were adapted for use in the calibration of these UVCANDELS data, including corrections for charge transfer efficiency according to \cite{Anderson2021}.   In addition, the images were individually corrected with custom hot pixel masks including readout cosmic rays, scattered light \citep[see][]{Biretta2003,Dulude2010}, and any 2D background gradient across the entire CCD chips.  The data were registered and stacked within a pipeline adapted from \cite{Alavi2014} using AstroDrizzle to combine the calibrated, flat fielded WFC3/UVIS and ACS/WFC images. Images were produced with 30 mas pixels and aligned to the CANDELS astrometric reference grid at a precision of 0.15 pixel, using unsaturated stars and compact sources. These UVCANDELS imaging data reach a depth of AB=27 (28) for compact galaxies in the WFC3/UVIS F275W (ACS/F435W). The final mosaiced data products are now available on MAST.\\ \vspace*{-15pt}


\begin{figure*}
\begin{center}
    \includegraphics[width=0.95\textwidth]{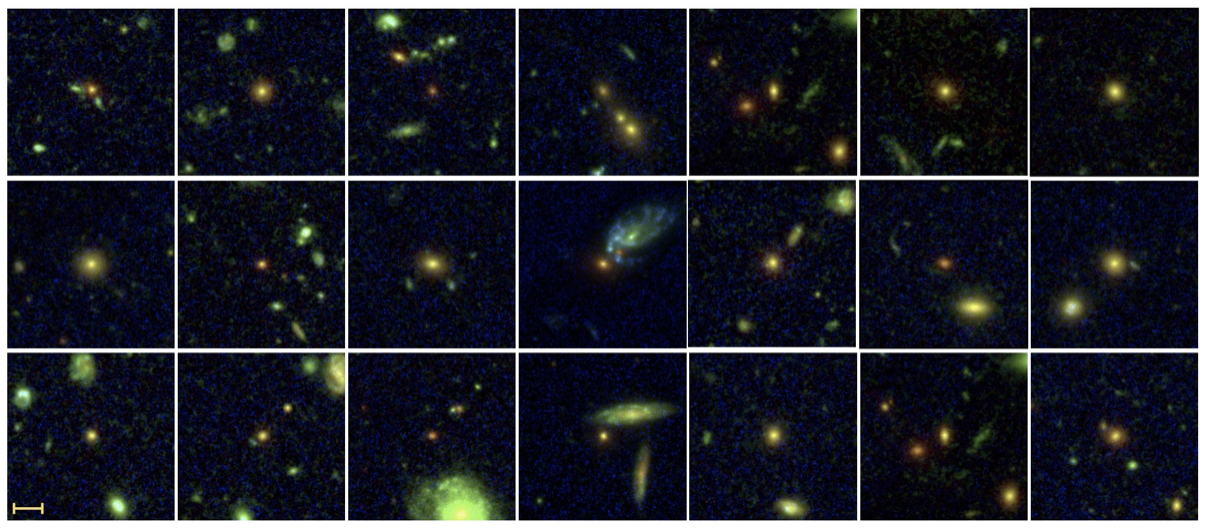}
    \caption{Representative three-color images of UVJ-selected, quiescent galaxies in the UVCANDELS fields considered in this survey. We combine UVCANDELS and archival imaging in WFC3/UVIS F275W, ACS/WFC F606W, and WFC3/IR F125W (B:G:R, respectively). The quiescent galaxy is centered in each image; a scalebar of length $\sim2\arcsec$ is provided (bottom left pane).}
        \label{fig:repgalaxies}
\end{center}
\end{figure*}

\section{Quiescent galaxies at $0.5 \le \lowercase{z} \le 1.5$}\label{sec:selection}

This survey used extensive, deep optical--near-IR imaging \& spectroscopy (m$\lesssim$26 AB) from MAST\footnote{\href{MAST}{https://archive.stsci.edu/prepds/3d-hst}}, including HST/ACS CANDELS broadband imaging \citep{Grogin2011, Koekemoer2011} and HST/IR G141 grism spectroscopy with ancillary catalogs \citep{Momcheva2016} which combined these grism data with extensive ground-based observations.\\ \vspace{-10pt}

\begin{figure*}
\begin{center}
    \includegraphics[width=0.95\textwidth]{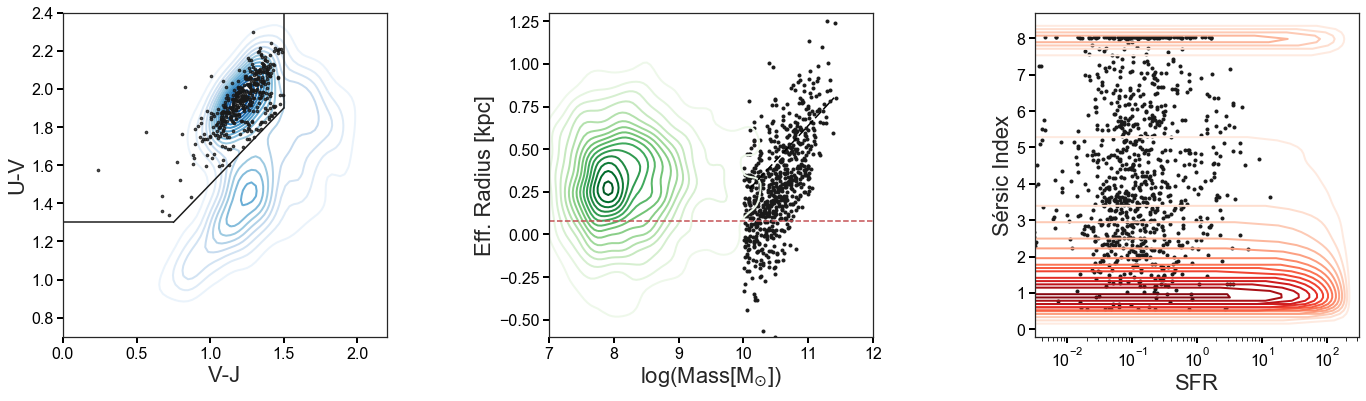}
    \caption{Relevant archival parameters derived for the UVCANDELS sample of quiescent galaxies; throughout, filled circles indicate significantly ($>3\sigma$) F275W-detected quiescent galaxies. {\it Left Panel}: UVJ color-color used for differentiating $0.5<z<1.5$ ``quiescent" (upper left) from SFGs; rest-frame UVJ colors from \cite{Momcheva2016}. {\it Center Panel}: Physical size [kpc]-stellar mass distribution [$M_{\odot}$] with the same CANDELS field galaxies (green contours), with the HST WFC3-IR/F160W PSF FWHM in physical units ($z\simeq1$). {\it Right Panel}: S\'{e}rsic profile index-star formation rate [$M_{\odot} yr^{-1}$] distribution with the same for CANDELS field galaxies (red contours). In Center \& Right panels, physical size parameters \cite{vanderwel2012} and stellar parameters are from \cite{Momcheva2016}.}\vspace*{-10pt}
   \label{fig:UVJselection}
\end{center}
\end{figure*}

Sample selection was made on well-determined massive ($M_{\star} \ge 10^{10}$ \Msol) galaxies at intermediate redshift ($0.5 \le z \le 1.5$) within the four UVCANDELS footprints, using 3DHST data products accessed via MAST. ext, using value-added catalogs of compiled photometry \citep{Momcheva2016}, we applied a rest-frame UVJ color-color criterion, and removed galaxies matched to X-ray detections within 0$\farcs$5 (83 total; priv. comm., D. Kocevski) which we assume to indicate AGN.  We identified a total of 1067 quiescent galaxies; false-color RGB images of a subsample of these galaxies are illustrated in Figure \ref{fig:repgalaxies}. We matched these galaxies on position, identifying matches within 0$\farcs$1 between the coaligned 3DHST and UVCANDELS catalogs. We define the statistics for the table in Table \ref{tab:galaxycount}. In Figure \ref{fig:UVJselection}, we present general physical characteristics of the full sample.


\section{Characterization of $0.5 \le \lowercase{z} \le 1.5$ Quiescent Galaxies}\label{sec:characterization}

\subsection{Quiescent galaxy morphologies}\label{subsec:galfit}

To assess quiescent galaxy morphology, we used \galfit\,\citep{Peng2010} to fit S\'{e}rsic models to CANDELS WFC3-IR/F125W 30mas imaging of the UVJ-selected sample. At 0.5$<z<$1.5, F125W is sensitive exclusively to the SED {\it redward} of the 4000\AA\ break and thus probes stellar emission from the old extant population. In this analysis, we enforced two constraints when implementing \galfit\,: 1) {\it Effective radius, $r_e$:} 0.5 pixels\lsim$r_e$\lsim100 pixels (or equivalently, $r_e\lesssim3\farcs5$); 2) {\it S\'{e}rsic index, $n$}: 0.5\lsim$n$\lsim8.
Here, the effective radius corresponds to the ``half-light'' radius, or the inner radius containing half of the total flux in the best-fitting S\'{e}rsic model fit.

The majority were well-fit ($\chi^2_{\nu}\lesssim5$), and we measured an average S\'{e}rsic index of $n\simeq4$ (to surface brightness $\mu_{F125W}\simeq29$mag sq.\! arcsec$^{-1}$), consistent with the spheroidal/bulge morphology expected for such galaxies.\\ \vspace*{-20pt}

\subsection{Size-mass distribution}\label{subsec:sizemass}

\begin{figure*}
\begin{center}
   \includegraphics[width=0.95\textwidth]{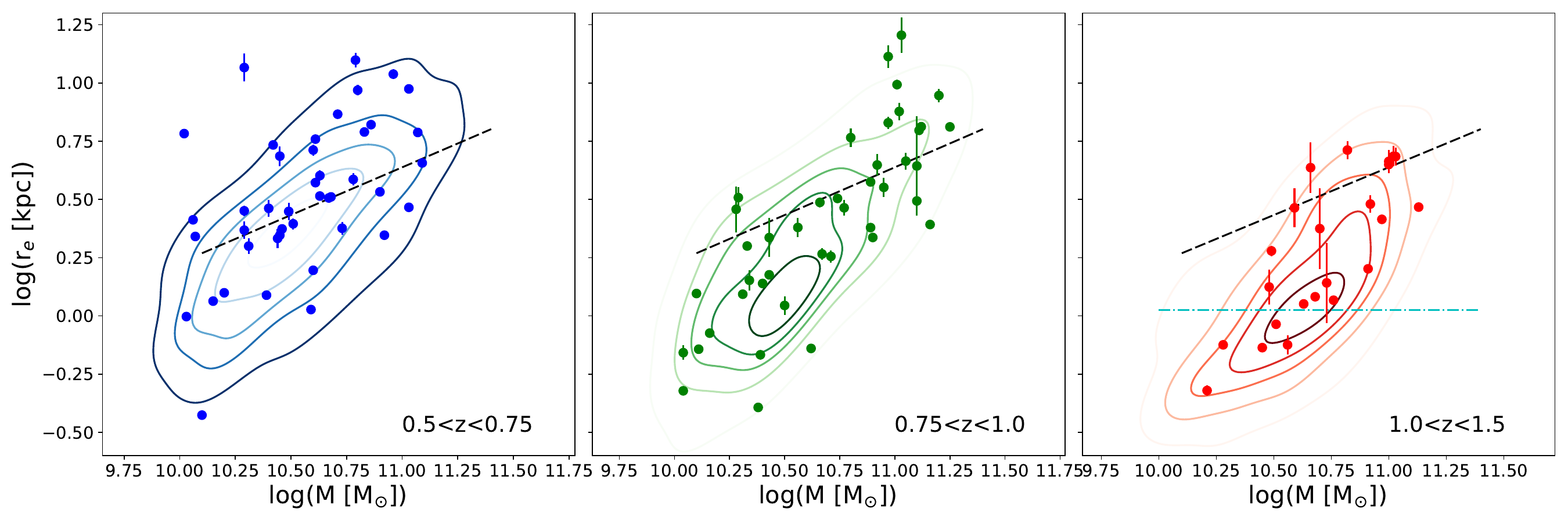}
     \caption{The effective (half-light) radius--stellar mass distribution measured for UVJ-selected galaxies in three redshift bins as indicated in each panel. Here, UVJ-selected, F275W-{\it detected} quiescent galaxies are indicated with filled points. Overplotted are density contours (over the range 0-100\%, in 20\% decrements) for size-mass distribution of the full UVJ-selected galaxies in the sample. For comparison, the local size-mass relation \citep[black, dashed;][]{Shen2003} is provided in each figure, with F125W FWHM in physical units (at $z\sim1$) is indicated in cyan in the Right panel.}
    \label{fig:sizemass} \vspace*{-15pt}
\end{center}
\end{figure*}

We provide context for the physical size-stellar mass distribution of the sample quiescent galaxies in Fig.~\ref{fig:sizemass}.  Here, for clarity, we include the galaxies well-fit\footnote{$1\lesssim \chi^2_{\nu}\lesssim 5$} with {\tt GALFIT}, separated into three redshift bins (equal in number).  Half-light stellar radii and stellar mass measurements for galaxies significantly ($>3\sigma$) detected in F275W at $0.5 \leq z < 0.75$ (blue; left panel), $0.75 \leq z < 1$ (green; center panel), and $1 \leq z \leq 1.5$ (red; right panel) are indicated by filled circles. We overplot size-mass measurements for all UVJ-selected quiescent galaxies in the same color scheme in the panels, with {\tt seaborn} kernel density estimates, as 20\% increments between 0 and 100\% density.

For ease of comparison, we define a convenience function, ``$\bar{x}_{r_e}$'' to compare the size-mass centroid in a given redshift bin (marginalized over mass) and the local \cite{Shen2003} relation (overplotted, black dashed). At $z<0.75$, the distribution of both the F275-detected and non-detected samples of UVJ-selected galaxies are consistent with the local relationship ($\bar{x}_{r_e}\lesssim1.2$; i.e., sizes are consistent within $\sim$20\%).  At $1 \leq z \leq 1.5$, though, galaxies are uniformly smaller ($\bar{x}_{r_e}\sim2$). Thus, consistent with previous work \citep{Mowla2019,Nedkova2021}, if these galaxies are analogs of progenitors to local quiescent galaxies, the class must undergo moderate size growth.    \\ \vspace*{-20pt}

\subsection{Recent star formation}\label{subsec:rsf}

Approximately 15\% of these UVJ-selected ``quiescent" galaxies were significantly detected in F275W ($\>3\sigma$) in UVCANDELS.  Recently-formed, massive stars are likely the source of this emission; our selection excludes X-ray detected galaxies that may host AGN, and these galaxies are too young cosmologically to have developed evolved stars \citep[e.g., hot horizontal branch or He-enhanced populations;][respectively]{Chung2011,Bekki2012} most likely responsible for ``UV Upturn" \citep{Yi1998}.

We investigated the observed UV-optical-nearIR SEDs of F275W-detected galaxies with {\tt CIGALE} \citep{Boquien2019}, assuming the galaxy redshift, $z_{best}$, and optical/nearIR observed photometry from \cite{Momcheva2016}.  In this analysis, we fit a library of $\sim10^7$ models defined by a composite stellar population defined with an ``Old" (consistent with formation redshift, $z_{form}$\gsim3) and a ``Young" ($\leq1$ Gyr\footnote{This ``young" age limit is set sufficiently high so as to include A-type stellar populations, for which ``E$+$A" galaxies are named\citep[see][]{Dressler1983}; the more common sobriquet ``post-starburst" encapsulates this class.}, with young-to-total stellar mass fraction, $f_m$,$\in\{10^{-2},10^{-1},1,10\}[\%]$) population. These models' SFHs were defined by either 1) the two-component exponentially-declining SFH ({\tt CIGALE:sfh2exp}) model or a 2) ``delayed" exponential-burst ({\tt CIGALE:sfhdelayed}) --- in both cases, ages of the young population were coarsely gridded, $t_{young}$[Myr]$\in\{1,10,50,100,500,1000\}$. Stellar populations were developed from \cite{Bruzual2003} models, assuming $Z\in\{Z_{\odot}$, $0.5Z_{\odot}\}$ and a \cite{Charlot2000} dust attenuation model with $A_V\in\{0.0,0.1,0.5,1.0,1.5,2.0,2.5\}$. 

Both {\tt sfhdelayed} and {\tt sfh2exp} SFHs were considered in this modeling, as a preference for either in the SED fitting results could constrain the source of RSF as quenched, post-starburst and rejuvenated quiescent galaxies have experienced fundamentally different recent SFHs. In recently-quenched galaxies, the rapid quenching and subsequent decline into quiescence observed thereafter can be well-modeled with truncated SFHs or composite models with multiple, exponentially-declining SFHs \citep{Suess2022}. Alternatively, though dry mergers likely predominate \citep{Lin2010}, if ``wet'' (gas-rich) mergers occur and promote new star formation within an extant quiescent system, this rejuvenated star formation caught ``in the act" may be readily discerned as a UV-luminous burst in the SED \citep{Rowlands2017}. 

In practice, these SFH classes could not be differentiated in this modeling.  The reduced $\chi^2$ values measured for the best-fit SED model with either SFH differed by $\lesssim 1\%$ on average for the full sample. Thus, we made no further interpretation  of these model fits for differentiating the potential SFHs. We will discuss extensions to this effort (\S\ref{sec:discussion}) to resolve this degeneracy with new data.

\begin{table*}
\begin{center}
\caption{Quiescent galaxy selection and derived samples}
\begin{tabular}{lccccr}
\hline \hline
{\it Selection Criteria} & EGS & COSMOS & GOODS-N & GOODS-S& {\bf Total} \\
\hline
UVJ-selected quiescent galaxies  & 312 & 413 & 259 & 83 & {\bf 1067} \\
WFC3/F275W detection ($>3\sigma$) & 37 & 38 & 48 & 18 &  {\bf 141} \\
$\hookrightarrow$ WFC3/G141 Emission Line (H$\alpha>5\sigma$) & 8 & 7 & 6 & 3 &  {\bf 24} \\
\hline \hline
    \end{tabular}%
\label{tab:galaxycount}
\end{center}
\noindent\makebox[\textwidth][c]{%
    \begin{minipage}{.8\textwidth}\small {\bf Notes}: Tabulated galaxies were selected with $M_{\star}>10^{10}$ \Msol\ at redshift $0.5 \le z \le 1.5$ in the footprint common to UVCANDELS \& 3DHST.  ``Quiescent" galaxies selected on UVJ colors following \cite{Williams2009} are provided in the first row here; in the second row, UV-detected, X-ray non-detections.  We note galaxies with significant H$\alpha$ emission ($5\sigma$), reported by \cite{Momcheva2016}. For further discussion of these samples, see Section \ref{sec:selection}.

            \end{minipage}}
    
\end{table*}


Nonetheless, this analysis provided important, novel constraints on RSF in these galaxies. The UVJ colors of quiescent galaxies typically arise from old, evolved stellar population in a galaxy with low specific SFR (sSFR), implying a passive evolution for an extended ($\gtrsim1Gyr$) timescale. Here, for the quiescent, F275W-detected objects, \cite{Mehta2024} in an independent SED analysis report a low average log(sSFR[yr$^{-1}$])$\simeq -11.4$.  We find a consistent low, average log(sSFR[yr$^{-1}$])$\simeq{-10.4}$ (comparable to the standard threshhold on sSFR for quiescent galaxies) while expanding parameter space to include an additional (young) model component. With {\tt CIGALE}, few ($<5\%$) of these ``quiescent" galaxies have SEDs consistent with {\it no} RSF, or stellar mass fraction strictly equal to zero. This result highlights the insensitivity of the UVJ selection criteria for differentiating quiescent galaxies with RSF from the general class of passively-evolving massive galaxies \citep{Leja2019}. Conversely, we confirm RSF for $>95\%$ of the F275W-detected galaxies well-fit ($\chi^2_{\nu}<5$) by these composite (young \& old) models using {\tt CIGALE}.

\begin{figure*}[htb]
\begin{center}
    \includegraphics[width=0.75\textwidth]{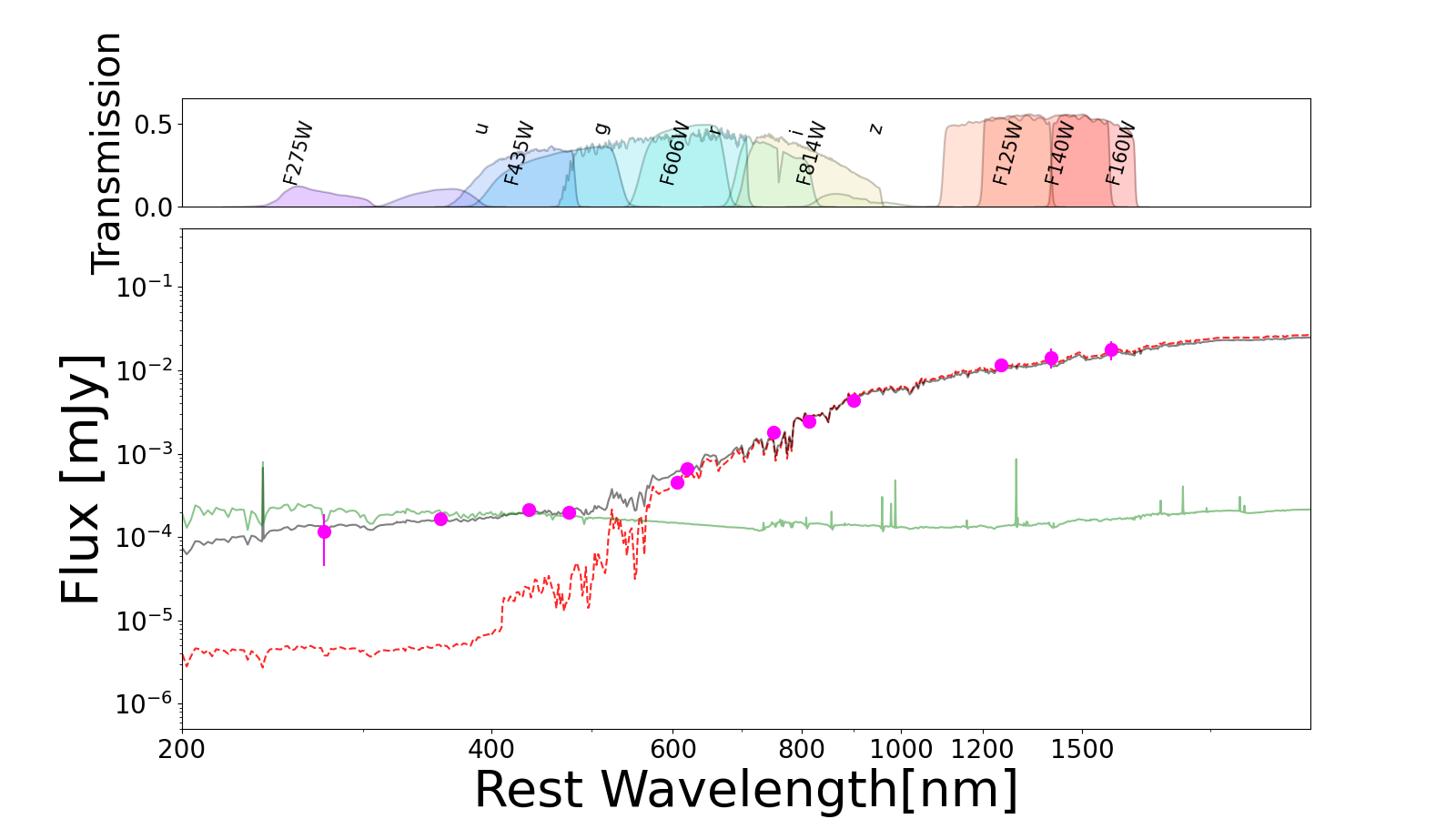}
    \caption{Overplotted on the SED (magenta, filled circles) of this representative UVJ-selected, F275W-detected quiescent galaxies are the best-fit young and old stellar populations (green and red, respectively) with the dust-corrected, best-fit model (black) population.  Without UVCANDELS F275W$+$F435W, the prominent very young ($t_{young}\simeq10$Myr), albeit minor ($f_m \simeq 0.01$\%), stellar population in this ``quiescent" galaxy is difficult to characterize. In the top panel, the broadband filtercurves used are provided.}
    \label{fig:sedgalaxy} \vspace*{-10pt}
\end{center}
\end{figure*}

\begin{figure*}[htb!]
\begin{center}
    \includegraphics[width=0.95\textwidth,trim=5in 5.5in 5.5in 3in]{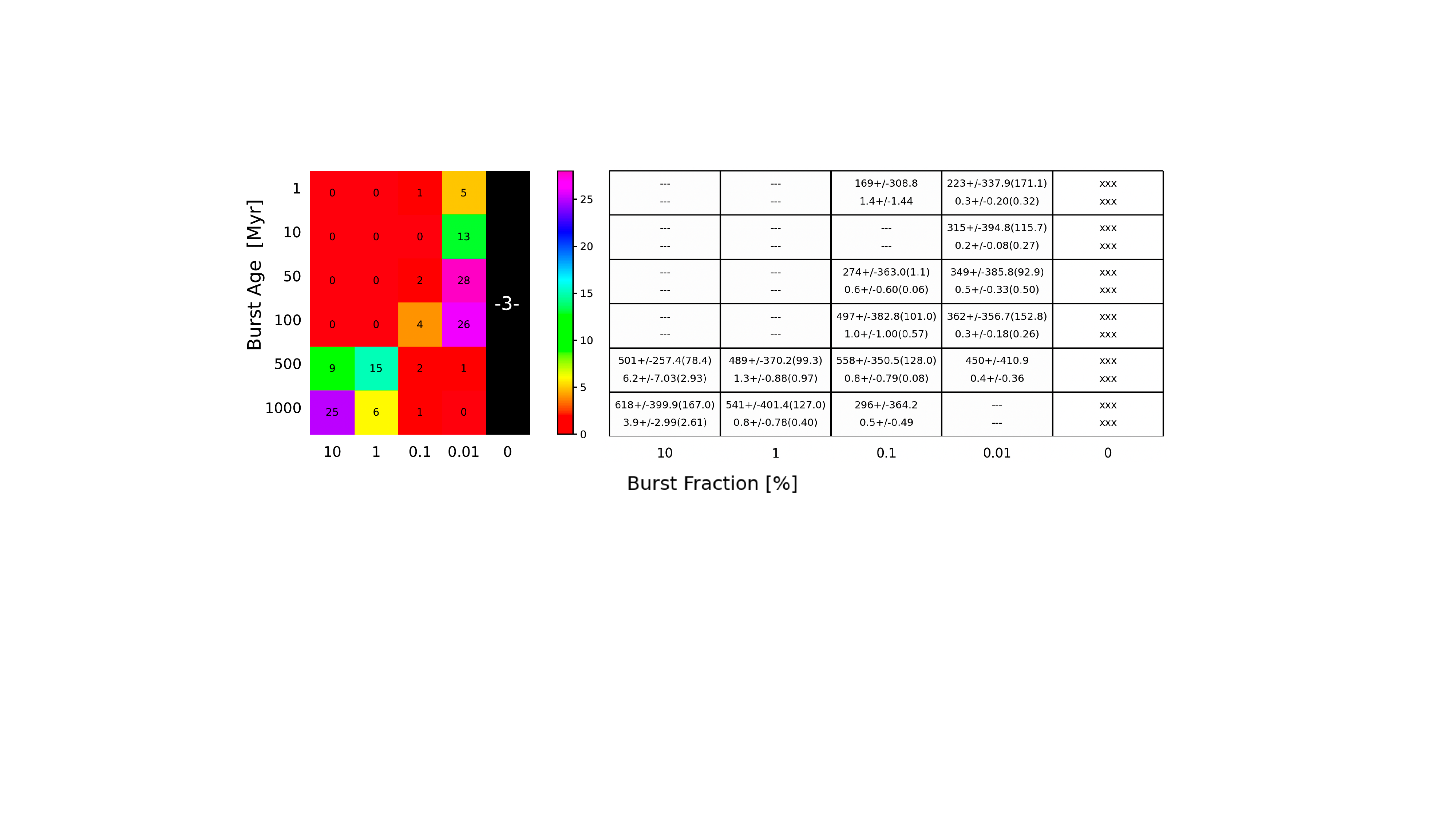}
    \caption{Young stellar age and stellar mass fraction derived with {\tt CIGALE} \citep[see \S\ref{sec:characterization} above, ][]{Boquien2019} for UVJ-selected, quiescent galaxies with significant F275W ($\geq3\sigma$) detections. {\it Left panel:} a two-dimensional heatmap histogram indicating {\tt CIGALE}-``best" parameter estimation.  {\it Right Panel:} confidence intervals derived from the {\tt CIGALE}-``Bayes" parameter estimation for all galaxies in each histogram cell. The upper (lower) row is formatted as follows as ``$x_0\pm x_1(x_2)$", with  $x_0$=mean, $x_1$= mean error, and $x_2$=standard deviation derived for the burst mass (fraction), respectively. Regions of parameter space not populated by galaxies in this sample are indicated ``---";  ``xxx" indicates  {\tt CIGALE} best-fit SED was inconsistent with young stellar population.  When a histogram cell is populated by a single galaxy, $x_2\equiv0.00$ is excluded.}
    \label{fig:twodhistmap} \vspace*{-15pt}
\end{center}
\end{figure*}

These UVCANDELS data can improve the age and mass fraction of the young stellar population associated with this RSF, as well. In Figure \ref{fig:sedgalaxy}, we plot the observed SED overplotted best-fit model derived with {\tt CIGALE} of a UV-detected quiescent galaxy, illustrating the utility of these UV data for deriving more robust constraints of the extant stellar populations in such galaxies. For the full sample of UV-detected quiescent galaxies, we summarize in Figure \ref{fig:twodhistmap} the best-fit model (i.e., lowest $\chi^2_{\nu}$) age and stellar mass fraction of the young stellar component, with confidence intervals for the best-fit models in each set derived from the {\tt CIGALE} Bayesian parameter estimations. With a small (2-3) number of independent, broadband rest-frame UV constraints on the SED, this model fitting marginally ($\sim2\sigma$) identifies RSF as associated with a young population of age, $t,\simeq200$Myr that is primarily distinguished by the stellar mass fraction.\\ \vspace*{-15pt}



\section{Local environments}\label{sec:environment}
Within hierarchical assembly, mergers are necessarily implicated in the formation and evolution of massive quiescent galaxies \citep{Bluck2012, Carleton2020,Conselice2022,Seuss2023}. A strong correlation of RSF with environment may suggest an enhanced role for such processes. Specifically for this sample, the availability of extensive photometric redshift catalogs (providing a coarse constraint on the density of galaxies in the local environment) makes it possible to meaningfully address the question: ``What is the probability that a quiescent galaxy will be identified with RSF, given its local environment?" A question of binary classification such as this, whether a galaxy is detected in F275W or not on the basis of a predictor factor (here, a general measure of environmental richness) is one for which a logistic regression statistical analysis is uniquely well-suited and appropriate. 

We investigated the potential for environmental effects to promote RSF in these quiescent galaxies. For this measurement, we first defined a uniform volume (spatially: $r\lesssim120$ kpc; $\theta<15$'' at $z\approx1$; in velocity: $\Delta z_{phot}<0.1$, or $\sim2x$ the mean uncertainty on redshift), for all UVJ-selected quiescent galaxies. Using the archival photometric redshift and image data from \cite{Momcheva2016}, $m_{606}\lesssim25$ objects in this volume were tabulated as ``Phot-Z Neighbors" to the quiescent galaxy; in velocity-space, neighbors were at the redshift of the quiescent galaxy, within $1\sigma$ of their measured photometric redshift uncertainty. This measure of ``neighbors" is fairly robust to the intrinsic variation by {\it increased} volume (increasing by $\sim\!3\times$) and {\it decreased} sensitivity (decreasing by $\sim\!9\times$\footnote{surveys with a fixed minimum observed magnitude, such as this one from which we take photometric redshifts, is inclusive of galaxies intrinsically $\sim9\times$ brighter at the minimum redshift extrema of the full sample}) over the redshift range surveyed.  For three equally-sized redshift bins,  we find a consistent median number of ``Phot-z Neighbors" equal to $5\pm3.5(1\sigma)$. 

Note, this general measure of environment defines a proxy for environmental richness in such a way that these neighbors will likely merger by $z\sim0$ \citep[see simulations by][]{Jian2012,Tal2013}.  We caution against direct comparison between the number of Phot-z Neighbors and other published classifications of group membership, though some may be broadly similar \citep[e.g., ``poor groups"][]{Poggianti2009}.

We developed a straightforward MCMC model which applies a logistic regression analysis to determine the extent to which the number of neighbors and detection of F275W emission of these quiescent galaxies are correlated. In this analysis, we assumed a sigmoidal distribution, the standard for logistic regression, defined as:\\ \vspace*{-20pt}

\begin{equation}
f(x)
 = \frac{1}{1+exp\{-((\beta_1 \times x)+ \beta_0)\}}, \\
\end{equation}
$f(x)$ is the probability of F275W detection and {\it $x$}, the predictor, is the number of Phot-z Neighbors. The terms $\beta_0$ and $\beta_1$ were defined as normal priors and fitted during the MCMC process: $\beta \sim Norm(\mu,\sigma)$, where $Norm$ is the univariate normal log-likelihood. Specifically, we employed non-informative priors defined by $\beta=Norm(0,10)$ such that the probability of detection be ultimately measured in this analysis was guided by the predictor itself rather than the choice of prior. We found our model to be relatively robust with varying these priors. Note, RSF is a stochastic, multi-variate process of finite duration, and its detection via F275W emission is necessarily dependent on the epoch at which the galaxy is observed.  The development and application of a complete, physically-motivated suite of priors is beyond the scope of this Letter. 

\begin{figure}[htb!]
\begin{center}
    \includegraphics[width=0.45\textwidth]{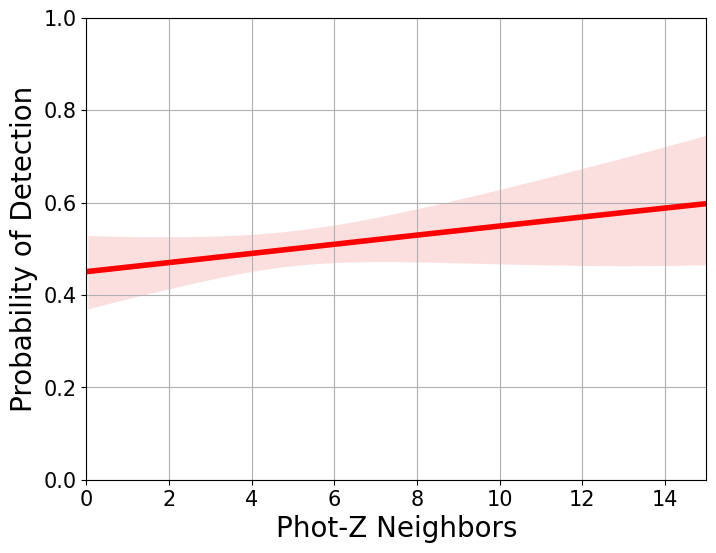}
    \caption{We implemented an MCMC logistic regression statistical analysis to determine the probability of a F275W detection for quiescent galaxies --- a signpost for RSF (\S\ref{subsec:rsf})---given the number of neighbors.   We find the probability of a detection increases marginally with the number of neighbors.  This weak environmental correlation likely indicates a reduced role for environmental processes (mergers, interactions), amongst the array of internal and ex-situ processes, in the formation and evolution of quiescent galaxies.}
    \label{fig:bootstrap} \vspace*{-20pt}
\end{center}
\end{figure}

Fundamentally, we used a standard MCMC approach to measure posteriors for each of the $\beta$ parameters \citep[implemented via {\tt pymc},][]{Salvatier2016}. This MCMC is implemented within a Monte Carlo wrapper though, which ameliorated the disparate sizes the F275W-detected and non-detected samples. Specifically, in each MC iteration, we defined the predictor array as the concatenation of the F275W-detected sample array and a random draw from the F275W non-detected quiescent galaxies (of the same length as the F275W-detected sample) with the draw weighted to ensure this sub-sample has a similar distribution of Phot-z Neighbors as the {\it full} F275W non-detected sample itself.  In each MC iteration, we calculated the predictor probabilities (i.e., applying the sigmoid),  the F275W detection likelihood was generated ({\tt pymc.Bernouilli}), and with the MCMC we then measured the $\beta$ parameters. We repeated the MC process 500 times, and recorded the Maximum A Posteriori (MAP) in each instance.  At completion, we averaged the MAP over the full MC run, and confirmed that the MAP does not change significantly (i.e., no flukes occurred in the sampling process). Applying this result, we represent the F275W detection probability in Figure \ref{fig:bootstrap}, with the 68\% maximum credible limits (determined with Python {\tt arviz.hdi}). This analysis yields a {\it marginal} positive correlation. Quiescent galaxies in richer (larger numbers of neighbors) environments may be more likely to experience RSF, but the role of the environment in the recent star formation in quenched or quiescent galaxies appears to be minor.

\section{Discussion}\label{sec:discussion}
Nearly all F275W-detected, UVJ-selected quiescent galaxies are consistent with recent SFH; i.e., $>$95\% of these galaxies are  best-fit with models for which the stellar mass fraction of the young stellar component is {\it strictly} models non-zero.  This fraction of all quiescent galaxies surveyed is lower ($\sim15\%$) than has been historically reported in the literature. Of that fraction, $\sim60\%$ of quiescent galaxies likely have experienced RSF within the past 100Myr but this young population constituted less $\sim1\%$ of the total stellar mass, albeit with considerable uncertainty. We note that only a small subset ($\sim1\%$) of these F275W-detected quiescent galaxies with confirmed RSF from SED modeling are identified with weak ($\sim3-5\sigma$) H$\alpha$ emission.  Thus, these HST UV data were critical for identifying and characterizing RSF in large area imaging surveys of quiescent galaxies evolution.

Mergers and interactions in particular have been targeted for studies to  characterize the frequency and dominance over cosmic time of the internal and ex-situ process driving quiescent galaxy evolution. In these efforts, morphological evidence of merger activity in, e.g., rejuvenating quiescent galaxies is rare in the local universe \citep{Crockett2012};  at low redshift morphological proxies for mergers \citep{Conselice2003} or correlations with, e.g., environmental density \citep[see, e.g.,][]{Cleland2021,Wilkinson2022} are used to infer the role for mergers. In hierarchical assembly, galaxy merger rates are expected to correlate with environment and increase with decreasing mass ratio of the mergers as demonstrated in cosmological simulations \citep{Fakhouri2009,Jian2012}. Note, recent observations at $z\lesssim0.3$ indicate, amongst field or group massive quiescent galaxies, the ensemble major merger fraction may in fact be weakly or {\it anti}-correlated with environmental richness \citep{Pearson2024, Sureshkumar2024}.
 
We found a {\it marginal} positive correlation between the number of neighbors and RSF.  If molecular gas accretion to massive, quiescent galaxies through mergers and interactions promoted the observed RSF \citep{Patton2020}, this weak correlation could suggest a minor role specifically for {\it gas-rich} (wet) accretion, supporting independent conclusions inferred from optical morphology \citep{Ji2022}.   Alternatively, RSF may arise in recently-quenched galaxies manifesting as PSBs.  Such galaxies constitute increasingly larger fractions of galaxies in increasingly denser environments \citep{Poggianti2009,Paccagnella2019}, but the marginal correlation suggests a minor role for mergers and by implication, a more pronounced role for other (e.g., AGN; X-ray {\it faint} given our selection) feedback modes \citep{Smethurst2016,MartinNavarro2022}. Note, few of the sample galaxies are spatially resolved in the UVCANDELS F275W images, but in future work we will investigate UV-optical color gradients, extending recent efforts at longer wavelengths \citep{Guo2011,Ji2023,Cheng2024} to characterize the assembly histories of the galaxies. This analysis could assist in revealing the extent to which mergers and interactions enhance the predominant smooth accretion mode identified by simulations \citep[see][]{Padmanabhan2020}. 

Finally, high resolution rest-frame optical-near IR spectra are important for advancing the effort presented in this Letter. First, spectroscopic redshifts improve the assessment of the environment (close pair and group identification and velocity dispersions therein) over what can be achieved with photometric and grism redshifts primarily relied upon in this work.  Furthermore, improved stellar mass and companion SFHs allows for the extensions of the logistic regression analysis in \S\ref{sec:environment} to additional predictors (e.g., quiescent to companion stellar mass ratios, useful for discriminating major and minor mergers). Secondly, high-resolution spectra are critical for identifying the specific RSF modality: rejuvenation or post-starburst. PSBs are often identified spectroscopically amongst quiescent galaxies on the basis of strong Balmer absorption\footnote{Though not standardized, PSBs are often selected on H$\delta$ absorption strength \citep[H$\delta>5$; e.g.][]{Goto2003,Alatalo2016}, with other constraints commonly employed \citep[e.g., (NUV-g$^{\prime}$) color,][]{Yesuf2017}} which are inaccessible to low-resolution HST grism spectroscopy for individual galaxies, and the truncated SFH of these galaxies is not readily distinguished by UV-optical-near IR broadband data alone \citep[see][]{Suess2022}.   Distinguishing PSBs from rejuvenated galaxies in this sample constrains the pathway to quiescence, which has broader applicability for questions of quiescent galaxy evolution.  For example, \cite{Zhang2024} recently selected PSBs from the Dark Energy Survey, and measured nearIR stellar sizes using HST.  They found that PSBs were typically smaller than the extant quiescent galaxy population. The authors suggest that (dry, minor) mergers could drive the size-mass growth of these galaxies. Determining the primary pathway by which progenitors of modern quiescent galaxies evolve from high redshift galaxies is made difficult by the selection of PSBs alone, which--as the authors state-- may be a biased progenitor set.  In this Letter, we show the utility of high spatial resolution imaging and rest-frame UV sensitivity for differentiating quiescent galaxies with and without RSF. When these data are combined with high-resolution spectroscopic data necessary for precise SFHs and more robust classification of the local environment, the mode(s) by which galaxies transform from star-forming to quiescent over cosmic time will be more readily constrained. 

\section{Conclusions}\label{sec:conclusions}

We have combined UVCANDELS F275W$+$F435W imaging with archival rest-frame optical-near IR photometry to investigate the class of UVJ color-color selected quiescent galaxies at \zrange. Applying this long baseline of wavelength coverage observed with HST over a decade, we determined $\sim15\%$ to have experienced RSF ($t<$1Gyr ago, with mass fraction $-1<$log($f_{M\star}$)$<-3$).  We performed a logistic regression statistical analysis to test for a correlation of of RSF with environmental richness. We found the RSF to be only marginally positively correlated with the environmental richness.  This correlation --- combined with the relatively small total mass in young stars --- may imply a weak role for the environmental processes in the evolution of this class of recently star-forming massive quiescent galaxies. Future efforts to further differentiate the modes of RSF will benefit from a high resolution spectroscopic campaign of the quiescent galaxies and their near-field companions.
\\ 

\section*{Acknowledgements}
This research is partly based on observations made with the NASA/ESA Hubble Space Telescope obtained from the Space Telescope Science Institute, which is operated by the Association of Universities for Research in Astronomy, Inc., under NASA contract NAS 5–26555. These observations are associated with program(s) HST-GO-15647. This research used archival data and value-added catalogs hosted by the Mikulski Archive for Space Telescopes at the Space Telescope Science Institute from the HST-GO-11600, HST-GO-12060, HST-GO-12099, HST-GO-12177, and HST-GO-12328 programs. We thank an anonymous referee whose insightful comments were helpful in the presentation of the data and analysis. MR acknowledges his late father, Melvin, for helpful discussions. X.W.~is supported by the National Natural Science Foundation of China (grant 12373009), the CAS Project for Young Scientists in Basic Research grant No.~YSBR-062, the
Fundamental Research Funds for the Central Universities, the
Xiaomi Young Talents Program, and the science research grant
from the China Manned Space Project. X.W.~also acknowledges
work carried out, in part, at the Swinburne University of
Technology, sponsored by the ACAMAR visiting fellowship.

\software{
\textsc{Astropy} \citep{astropy2013,astropy2022},
\textsc{Galfit} \citep{Peng2002,Peng2010},
\textsc{Cigale} \citep{Boquien2019}
\textsc{matplotlib} \citep{matplotlib2007}
\textsc{seaborn} \citep{seaborn}
}
\facilities{HST (WFC3), MAST (HLSP)}

\bibliographystyle{apj} 
\bibliography{UVJbibliography} 

\end{document}